# Frequency characteristics based on describing function method for differentiators

Xinhua Wang

Advanced Control Technology, Department of Electrical & Computer Engineering,

National University of Singapore, Singapore 117576

Email: wangxinhua04@gmail.com

**Abstract:** In this paper, describing function method is used to analyze the characteristics and parameters selection of differentiators. Nonlinear differentiator is an effective compensation to linear differentiator, and hybrid differentiator consisting of linear and nonlinear parts is the combination of both advantages of linear and nonlinear differentiators. The merits of the hybrid differentiator include its simplicity, rapid convergence at all times, and restraining noises effectively. The methods are confirmed by some examples.

## 1. Introduction

Differentiation of signals is a well-known problem [1-6], which has attracted much attention in recent years. Obtaining the velocities of tracked targets is crucial for several kinds of systems with correct and timely performances, such as the missile-interception systems [7] and underwater vehicle systems [8], in which disturbances must be restrained. The convergence rates, computation time and robustness of differentiators also should be taken into consideration.

The popular high-gain differentiators [4, 19, 22] provide for an exact derivative when their gains tend to infinity. Unfortunately, their sensitivity to small high-frequency noise also infinitely grows. With any finite gain values such a differentiator has also a finite bandwidth. Thus, being not exact, it is, at the same time, insensitive with respect to high-frequency noise. Such insensitivity may be considered both as advantage or disadvantage depending on the circumstances. In fact, this differentiator has the ability of restraining noises to some extent.

In [5, 6], a differentiator via second-order (or high-order) sliding modes algorithm has been proposed. The information one needs to know on the signal is an upper bound for Lipschitz constant of the derivative of the signal. This constrains the types of input signals. For this differentiator, the chattering phenomenon is inevitable.

In [18], we presented a finite-time-convergent differentiator based on finite-time stability [15-17] and singular perturbation technique [9-14]. The merits of this differentiator exist in three aspects: rapidly finite-time convergence compared with other typical differentiators; no chattering phenomenon; and besides the derivatives of the derivable signals, the generalized derivatives of some classes of signals can be obtained. However, the differentiators in [18] are complicated and difficult to be implemented in practice, due to the long computation time.

In [20], we designed a hybrid differentiator with high speed convergence, and it succeeds in application to velocity estimation for low-speed regions only based on position measurements [21], in which only the convergence of signal tracking was described for this differentiator, but the convergence of derivative tracking was not given, and the regulation of parameters has no rules.

For nonlinear and hybrid differentiators, an extended version of the frequency response method, describing function method [23, 24], can be used to approximately analyze and predict nonlinear behaviors of nonlinear and hybrid differentiators. Even though it is only an approximation method, the desirable properties it inherits from the frequency response method, and the shortage of other, systematic tools for nonlinear differentiator analysis, make it an indispensable component of the bag of tools of practicing control engineers. By describing function method, we can obtain the regulation of parameters of differentiators to restrain noises effectively and obtain the derivative of signal. Nonlinear differentiator is an effective compensation to linear differentiator, and hybrid differentiator consisting of linear and nonlinear parts is the combination of both advantages of linear and nonlinear differentiators.

## 2. Frequency domain Analysis of differentiators
### 2.1 Frequency Analysis of Linear differentiator

The linear differentiator is designed as [19, 20]:

$$\begin{aligned}\dot{x}_1 &= x_2 \\ \varepsilon^2 \dot{x}_2 &= -a_0(x_1 - v(t)) - a_1 \varepsilon x_2 \\ y &= x_2\end{aligned} \quad (1)$$

For derivable signal $v(t)$, there exist the relations as follows:

$$x_2 \to v'(t) \text{ as } \varepsilon \text{ is sufficiently small.}$$

In fact, linear differentiator (1) is equivalent to the one presented in [4, 22]. Define

$$x_1 = w_1 - \varepsilon a_1 w_2/a_0, \quad x_2 = w_2 \quad (2)$$

we have

$$\begin{aligned}\dot{w}_1 &= w_2 - a_1(w_1 - v(t))/\varepsilon \\ \dot{w}_2 &= -a_0(w_1 - v(t))/\varepsilon^2 \\ y &= w_2\end{aligned} \quad (2)$$

For linear differentiator (1), let $x = x_1, \dot{x} = x_2$, we have

$$\ddot{x} + \frac{a_0}{\varepsilon^2}(x - v(t)) + \frac{b_0}{\varepsilon}\dot{x} = 0 \quad (3)$$

We can transfer (3) to the following structure

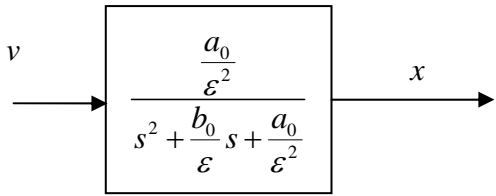

Fig. 1 The structure of transfer function of linear differentiator

Let $\omega_n^2 = a_0/\varepsilon^2$ therefore, we obtain the natural frequency as follow:

$$\omega_n = \sqrt{a_0}/\varepsilon, \quad (4)$$

and let $\dfrac{b_0}{\varepsilon} = 2\varsigma\omega_n$, and $0 < \varsigma < 1, \omega_n > 0$, therefore,

$b_0/\varepsilon = 2\varsigma\sqrt{a_0}/\varepsilon$, therefore, we get the damping coefficient

$$\varsigma = b_0/(2\sqrt{a_0}) \quad (5)$$

The closed-loop transfer function of linear differentiator (1) is:

$$G(s) = \frac{1}{(s/\omega_n)^2 + 2\varsigma(s/\omega_n) + 1} \quad (6)$$

and its frequency characteristic is:

$$G(j\omega) = \frac{\omega_n^2}{(s + \varsigma\omega_n + j\omega_d)(s + \varsigma\omega_n - j\omega_d)}\bigg|_{s=j\omega} \quad (7)$$

where

$$\omega_d = \omega_n\sqrt{1-\varsigma^2} \quad (8)$$

Amplitude-frequency characteristics is:

$$|G(j\omega)| = \frac{1}{\sqrt{(1-\omega^2/\omega_n^2)^2 + 4\varsigma^2\omega^2/\omega_n^2}} \quad (9)$$

and phase-frequency characteristics is:

$$\angle G(j\omega) = \begin{cases} -arctg\dfrac{2\varsigma\omega/\omega_n}{1-\omega^2/\omega_n^2}, & \text{when } \omega/\omega_n \le 1 \\ -\left(\pi - arctg\dfrac{2\varsigma\omega/\omega_n}{\omega^2/\omega_n^2 - 1}\right), & \text{when } \omega/\omega_n > 1 \end{cases} \quad (10)$$

In order to restrain disturbances, a suitable $\omega_n$ should be required. Therefore, from (4), $a_0$ and $\varepsilon$ decides $\omega_n$. $\varepsilon$ should not be selected too small in order to peaking phenomena happen, and $a_0$ should not be too big to make too much noises pass through the differentiator. From (10), the bigger the damping coefficient $\varsigma$ is, the bigger the phase shift is. Therefore, from (5), $b_0$ cannot be selected too big.

Log magnitude-frequency characteristics of oscillation element are:

$$L(\omega) = -20\lg\sqrt{(1-\omega^2/\omega_n^2)^2 + (2\varsigma\omega/\omega_n)^2} \quad (11)$$

When $\omega \ll \omega_n$, i.e., $\omega \ll \sqrt{a_0}/\varepsilon$,

$$L(\omega) \approx 0 \quad (12)$$

When $\omega \gg \omega_n$, i.e., $\omega \gg \dfrac{\sqrt{a_0}}{\varepsilon}$,

$$L(\omega) \approx -40\lg(\omega/\omega_n) \quad (13)$$

In the following, we compare the filter characteristics of linear differentiator with classical first-order filter. We know the transfer function of the classical first-order filer is:

$$G(s) = \dfrac{\sqrt{a_0}/\varepsilon}{s + \sqrt{a_0}/\varepsilon} \quad (14)$$

Amplitude-frequency characteristics is:

$$|G(j\omega)| = \dfrac{1}{\sqrt{\omega^2 \varepsilon^2/a_0 + 1}} \quad (15)$$

and phase-frequency characteristics:

$$\angle G(j\omega) = -\arctg\left(\omega\varepsilon/\sqrt{a_0}\right) \quad (16)$$

Log magnitude-frequency characteristics of oscillation element are:

$$L(\omega) = -20\lg\sqrt{\omega^2\varepsilon^2/a_0 + 1} \quad (17)$$

When $\omega \ll \sqrt{a_0}/\varepsilon$, we have

$$L(\omega) \approx -20\lg 1 = 0 \quad (18)$$

When $\omega \ll \dfrac{\sqrt{a_0}}{\varepsilon}$, we have

$$L(\omega) \approx -20\lg\left(\omega\varepsilon/\sqrt{a_0}\right) \quad (19)$$

From (13) and (19), in the high-frequency phase, linear differentiator has better ability of restraining disturbances comparing with the classical first-order filter.

## 2.2 Describing Function Analysis of Nonlinear differentiator

**Denotations:** $sig(y)^\alpha = |y|^\alpha \, \text{sgn}(y)$, $\alpha > 0$. It is obvious that $sig(y)^\alpha = y^\alpha$ only if $\alpha = q/p$, where $p, q$ are positive odd numbers. Moreover,

$$\dfrac{d}{dy}|y|^{\alpha+1} = (\alpha+1)sig(y)^\alpha, \quad \dfrac{d}{dy}sig(y)^{\alpha+1} = (\alpha+1)|y|^\alpha$$

For nonlinear differentiator [20]

$$\begin{aligned}\dot{x}_1 &= x_2 \\ \dot{x}_2 &= -\dfrac{a_1}{\varepsilon^2} sig(x_1 - v(t))^\alpha - \dfrac{b_1}{\varepsilon^2} sig(\varepsilon x_2)^\alpha\end{aligned} \quad (20)$$

where $a_1, b_1 > 0$, $0 < \alpha < 1$, and $\varepsilon > 0$ is a small perturbation parameter. From [18] and [20], we have that:

For a continuous two-order derivable signal $v(t)$, there exists $\gamma > 0$ (where $\alpha\gamma > 2$) and $\Gamma > 0$ such that

$$x_1 - v(t) = O(\varepsilon^{\alpha\gamma}), \quad x_2 - \dot{v}(t) = O(\varepsilon^{\alpha\gamma-1}) \quad (21)$$

for $t > \varepsilon\Gamma$. $O(\varepsilon^{\alpha\gamma-1})$ denotes the approximation of $\varepsilon^{\alpha\gamma-1}$ order between $x_2$ and $\dot{v}(t)$.

Let us assume that there is a limit cycle in the system and the oscillation signal $x$ is in the form of

$$x(t) = A\sin\omega t, \quad (22)$$

with $A$ being the limit cycle amplitude and $\omega$ being the frequency. We have the describing function for $sig(x)^\alpha$ as follow:

$$\begin{aligned}N(A) &= \dfrac{\dfrac{2}{\pi}\int_0^\pi sig(A\sin\omega t)^\alpha \sin\omega t\, d\omega t}{A} \\ &= \dfrac{2}{\pi}A^{\alpha-1}\Omega(\alpha)\end{aligned} \quad (23)$$

Because

$$\dfrac{2}{\pi}\int_0^\pi (\sin\omega t)^2 d\omega t = 1 \quad (24)$$

and $|\sin\omega t| \leq 1$, we have

$$1 < \dfrac{2}{\pi}\int_0^\pi |\sin\omega t|^{\alpha+1} d\omega t = \dfrac{2}{\pi}\Omega(\alpha) < 2 \quad (25)$$

for $\alpha \in (0,1)$.

Therefore, the approximated linear differentiator of nonlinear differentiator (20) is:

$$\dot{x}_1 = x_2$$
$$\dot{x}_2 = -\frac{a_1}{\varepsilon^2} A^{\alpha-1} \frac{2}{\pi} \Omega(\alpha)(x_1 - v(t))$$
$$- \frac{b_1}{\varepsilon^2} A^{\alpha-1} \frac{2}{\pi} \Omega(\alpha)(\varepsilon x_2)$$
(26)

Let $x = x_1, \dot{x} = x_2$, we have

$$\ddot{x} + \frac{b_1}{\varepsilon} A^{\alpha-1} \frac{2}{\pi} \Omega(\alpha) \dot{x}$$
$$+ \frac{a_1}{\varepsilon^2} A^{\alpha-1} \frac{2}{\pi} \Omega(\alpha)(x - v(t)) = 0$$
(27)

We can transfer (27) to the following structure:

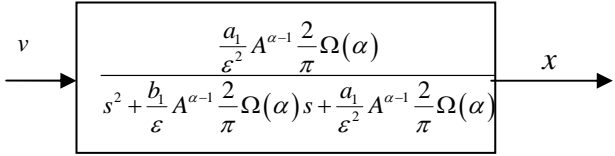

Fig. 2  The structure of approximated transfer function of nonlinear differentiator

Let $\omega_n^2 = \frac{a_1}{\varepsilon^2} A^{\alpha-1} \frac{2}{\pi} \Omega(\alpha)$, we have the natural frequency as:

$$\omega_n = \frac{\sqrt{a_1}}{\varepsilon} \sqrt{A^{\alpha-1} \frac{2}{\pi} \Omega(\alpha)},$$
(28)

and let $\frac{b_1}{\varepsilon} A^{\alpha-1} \frac{2}{\pi} \Omega(\alpha) = 2\varsigma \omega_n$, and $0 < \varsigma < 1$, $\omega_n > 0$, therefore, we get the damping coefficient as:

$$\varsigma = \frac{b_1 \sqrt{A^{\alpha-1} \frac{2}{\pi} \Omega(\alpha)}}{2\sqrt{a_1}}$$
(29)

Amplitude-frequency, log magnitude-frequency and phase-frequency characteristics are (9), (10) and (11), respectively.

When $\omega \ll \omega_n$, i.e., $\omega \ll \frac{\sqrt{a_1}}{\varepsilon} \sqrt{A^{\alpha-1} \frac{2}{\pi} \Omega(\alpha)}$, we have

$$L(\omega) \approx 0$$
(30)

When $\omega \gg \omega_n$, i.e., $\omega \gg \frac{\sqrt{a_1}}{\varepsilon} \sqrt{A^{\alpha-1} \frac{2}{\pi} \Omega(\alpha)}$, we get

$$L(\omega) \approx -40 \lg(\omega/\omega_n)$$
(31)

We can see that natural frequency of nonlinear differentiator become small as the amplitude of input signal turn big. Therefore, the nonlinear differentiator is suitable to compensate the filter of linear differentiator. Moreover, the noise with big amplitude can be restrained in low-frequency phase.

### 2.3 Describing Function Analysis of hybrid differentiator

In linear differentiator, if $a_0$ is selected as a relatively big value and $\varepsilon$ a small value, $\omega_n$ will be very big. This can make noise pass through the differentiator. If $\varepsilon$ is too small, the phenomenon happens. Therefore, $a_0$ should be selected as a relatively small value and $\varepsilon$ a relatively suitable value, this can make the output signal deformation and lag phenomenon happens. From the analysis of nonlinear differentiator, we can find that $\omega_n$ becomes smaller as the amplitude $A$ of input signal turns bigger. Accordingly, we firstly select $a_0$ as a relatively small value and $\varepsilon$ a relatively suitable value, and a nonlinear differentiator is introduced into the linear differentiator to compensate the low natural frequency of linear differentiator. When the amplitude of input signal is relatively mall, the natural frequency $\omega_n$ should be relatively big, and when the amplitude of input signal is relatively big, the natural frequency $\omega_n$ should keep relatively small.

For hybrid differentiator [20]
$$\dot{x}_1 = x_2$$
$$\varepsilon^2 \dot{x}_2 = -a_0(x_1 - v(t)) - a_1 sig(x_1 - v(t))^\alpha$$
$$- b_0 \varepsilon x_2 - b_1 sig(\varepsilon x_2)^\alpha$$
$$y = x_2$$
(32)

where $a_0, a_1, b_0, b_1 > 0$, $0 < \alpha < 1$, and $\varepsilon > 0$ is

a small perturbation parameter. From [18] and [21], we have that:

For a continuous two-order derivable signal $v(t)$, there exists $\gamma > 0$ (where $\alpha\gamma > 2$) and $\Gamma > 0$ such that

$$x_1 - v(t) = O(\varepsilon^{\alpha\gamma}), \quad x_2 - \dot{v}(t) = O(\varepsilon^{\alpha\gamma-1}) \quad (33)$$

for $t > \varepsilon\Gamma$.

The linear approximation of hybrid differentiator (32) by describing function method is:

$$\begin{aligned}
\dot{x}_1 &= x_2 \\
\dot{x}_2 &= -\frac{a_1}{\varepsilon^2} A^{\alpha-1} \frac{2}{\pi} \Omega(\alpha)(x_1 - v(t)) \\
&\quad - \frac{b_1}{\varepsilon^2} A^{\alpha-1} \frac{2}{\pi} \Omega(\alpha)(\varepsilon x_2) \\
&\quad - \frac{a_0}{\varepsilon^2}(x_1 - v(t)) - \frac{b_0}{\varepsilon} x_2
\end{aligned} \quad (34)$$

Therefore, we have

$$\begin{aligned}
\dot{x}_1 &= x_2 \\
\dot{x}_2 &= -\frac{1}{\varepsilon^2}\left(a_0 + a_1 A^{\alpha-1}\frac{2}{\pi}\Omega(\alpha)\right)(x_1 - v(t)) \\
&\quad - \frac{1}{\varepsilon^2}\left(b_0 + b_1 A^{\alpha-1}\frac{2}{\pi}\Omega(\alpha)\right)(\varepsilon x_2)
\end{aligned} \quad (35)$$

Let $x = x_1, \dot{x} = x_2$, we have

$$\ddot{x} + \frac{1}{\varepsilon}\left(b_0 + b_1 A^{\alpha-1}\frac{2}{\pi}\Omega(\alpha)\right)\dot{x} \\
+ \frac{1}{\varepsilon^2}\left(a_0 + a_1 A^{\alpha-1}\frac{2}{\pi}\Omega(\alpha)\right)(x - v(t)) = 0 \quad (36)$$

Let $\omega_n^2 = \frac{1}{\varepsilon^2}\left(a_0 + a_1 A^{\alpha-1}\frac{2}{\pi}\Omega(\alpha)\right)$, therefore, we have the natural frequency as:

$$\omega_n = \frac{1}{\varepsilon}\sqrt{a_0 + a_1 A^{\alpha-1}\frac{2}{\pi}\Omega(\alpha)}, \quad (37)$$

and let $\frac{1}{\varepsilon}\left(b_0 + b_1 A^{\alpha-1}\frac{2}{\pi}\Omega(\alpha)\right) = 2\varsigma\omega_n$, and

$0 < \varsigma < 1, \omega_n > 0$, therefore, we get damping coefficient as:

$$\varsigma = \frac{b_0 + b_1 A^{\alpha-1}\frac{2}{\pi}\Omega(\alpha)}{2\sqrt{a_0 + a_1 A^{\alpha-1}\frac{2}{\pi}\Omega(\alpha)}} \quad (38)$$

When the amplitude of input signal becomes big, this hybrid differentiator can still guarantee natural frequency by setting the suitable $a_0$ and $b_0$.

In the following, we give some examples to explain the frequency domain characteristics of linear, nonlinear and hybrid differentiators, respectively, and analysis of filtering abilities is discussed.

## 3. Examples for describing function analysis of linear, nonlinear and hybrid differentiators.

### A. *Linear differentiator (1):*

Parameters: $\varepsilon = \frac{1}{45}, a_0 = 0.05, b_0 = 0.3$

The transfer function is

$$G(s) = \frac{\frac{a_0}{\varepsilon^2}}{s^2 + \frac{b_0}{\varepsilon}s + \frac{a_0}{\varepsilon^2}} = \frac{101.25}{s^2 + 13.5s + 101.25}$$

The bode plots of linear differentiator are in the following:

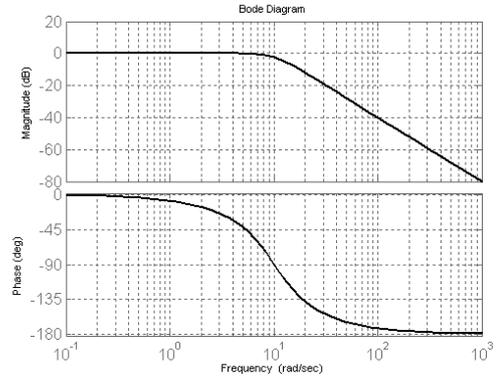

Fig. 3　Bode plots of linear differentiator

The natural frequency is

$$\omega_n = \frac{\sqrt{a_0}}{\varepsilon} = \frac{\sqrt{0.05}}{1/45} = 10.062$$

and damping coefficient is

$$\varsigma = \frac{b_0}{2\sqrt{a_0}} = \frac{0.3}{2\sqrt{0.05}} = 0.67$$

Input signal with noise:

5sin(2t)+Noise

Noise: Band-limited white noise, and noise power [0.01], sample time: 0.01.

The input signal with noise is in the following figure.

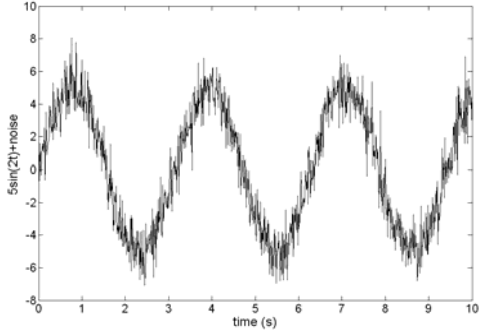

Fig. 4   Input signal 5sin2t with noises

The derivative output by linear differentiator is shown in the following figure:

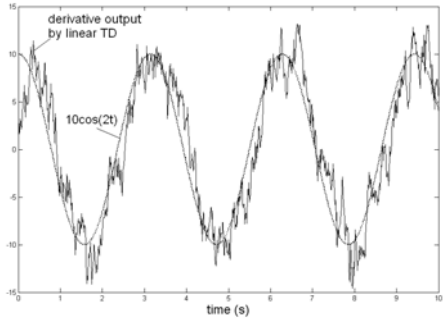

Fig. 5   The derivative output by linear differentiator

Because the natural frequency $\omega_n$ of linear differentiator is selected as a relatively big value, the noise also passes through the linear differentiator, and the effect of noise suppression is not satisfying.

*B.   Nonlinear differentiator (20):*

Parameters:

$$\varepsilon = \frac{1}{45}, a_1 = 0.099, b_1 = 0.268, \alpha = 0.5$$

We know that $A=5$, therefore, we have

$$G(s) = \frac{99.768}{s^2 + 6s + 99.768}$$

The bode plots of linear approximation of nonlinear differentiator are in the following:

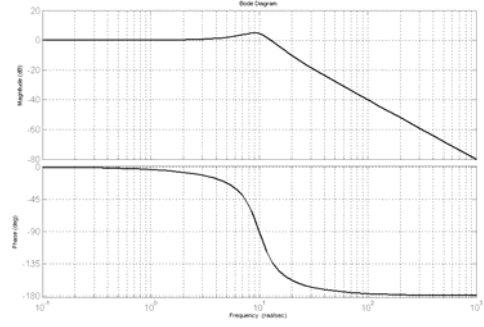

Fig. 6   Bode plots of linear approximation of nonlinear differentiator

The natural frequency is

$$\omega_n = \frac{\sqrt{a_1}}{\varepsilon}\sqrt{A^{\alpha-1}\frac{2}{\pi}\Omega(\alpha)} = \frac{\sqrt{0.099}}{1/45}\sqrt{1.1128 A^{-0.5}} = 10$$

and damping coefficient is

$$\varsigma = \frac{b_1\sqrt{A^{\alpha-1}\frac{2}{\pi}\Omega(\alpha)}}{2\sqrt{a_1}} = \frac{0.268\sqrt{1.1128 A^{-0.5}}}{2\sqrt{0.099}} = 0.3$$

We can find that the bigger the amplitude of input signal is, the smaller the natural frequency $\omega_n$ is.

The derivative output by nonlinear differentiator is shown in the following figure:

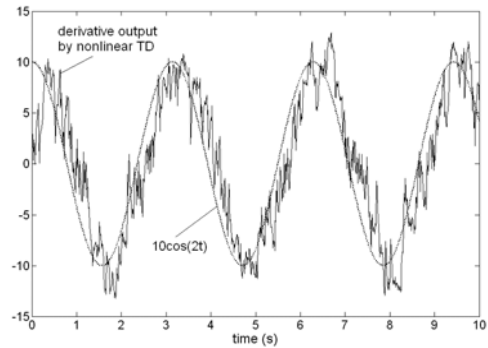

Fig. 7   The derivative output by nonlinear differentiator

*C. Noise restraining sufficiently by hybrid differentiator*

In order to restrain noises effectivlely, the natural frequency $\omega_n$ of linear differentiator should be selected as a relatively small value.

Input signal: 0.5sin2t

Noise: Band-limited white noise, and noise power 0.0001, sample time: 0.01.

The input signal with noise is in the following figure.

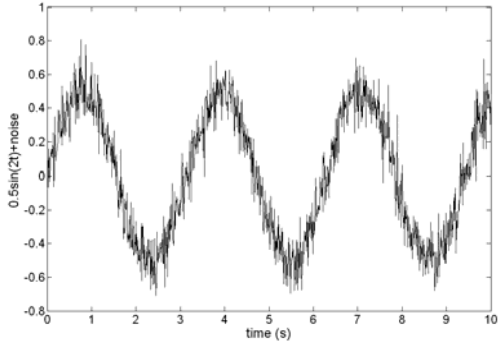

Fig. 8  Input signal 05sin2t with noises

Linear differentiator:

Parameters: $\varepsilon = 1/45, a_0 = 0.005, b_0 = 0.05$

The natural frequency is $\omega_n = 3.18$; and damping coefficient is $\varsigma = 0.35$.

Based on the design of linear differentiator, we given the parameters of hybrid differentiator in the following.

Hybrid differentiator:

Parameters:

$$\varepsilon = \frac{1}{45}, a_0 = 0.005, b_0 = 0.05, a_1 = 0.005,$$
$$b_1 = 0.005, \alpha = 0.5$$

The transfer function of linear approximated system by describing function is

$$G(s) = \frac{\frac{1}{\varepsilon^2}\left(a_0 + a_1 A^{\alpha-1}\frac{2}{\pi}\Omega(\alpha)\right)}{s^2 + \frac{1}{\varepsilon}\left(b_0 + b_1 A^{\alpha-1}\frac{2}{\pi}\Omega(\alpha)\right)s + \frac{1}{\varepsilon^2}\left(a_0 + a_1 A^{\alpha-1}\frac{2}{\pi}\Omega(\alpha)\right)}$$
$$= \frac{26.06}{s^2 + 2.6s + 26.06}$$

The curves of derivative outputs by linear and hybrid differentiators, respectively, are shown in the following figure.

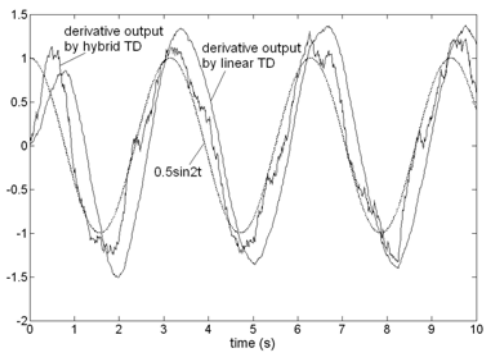

Fig. 9  Derivative outputs of linear and hybrid differentiators respectively

The natural frequency $\omega_n$ is selected as a relatively small value, the nonlinear parts compensate the natural frequency $\omega_n$ according to the effect of noises.

## 4. Frequency analysis of differentiators by sweept frequency

The frequency characteristics of the differentiators are identified by applying the sinusoidal inputs and measuring the amplitude and the phase lag of the output sinusoidal response with the same frequency. We denote: Linear TD (Linear tracking-differentiator, equation (1)), Nonlinear TD (Nonlinear tracking-differentiator, equation (20)), Hybrid TD (Hybrid differentiator, equation (32)), Low pass filter, Levant TD (Levant differentiator [5])

Let the input command is

$$v(t) = A\sin(\omega t)$$

The signal is assumed to have the frequency spectrum up to 100 Hz. The frequency characteristics of the differentiators with satisfying parameters

Linear TD: $1/\varepsilon = R = 100, a_0 = 0.1, b_0 = 0.3$

Nonlinear TD:

$1/\varepsilon = R = 45, a_1 = 0.015, b_1 = 0.015, \alpha_1 = \alpha_2 = 0.6$

Hybrid TD:

$$1/\varepsilon = R = 100, a_0 = 0.1, a_1 = 0.015, b_0 = 0.3,$$
$$b_1 = 0.015, \alpha_1 = \alpha_2 = 0.6$$

are identified by simulation. The frequency characteristics for the sinusoidal signals with fixed amplitude $A=1$ are shown in figure 10.

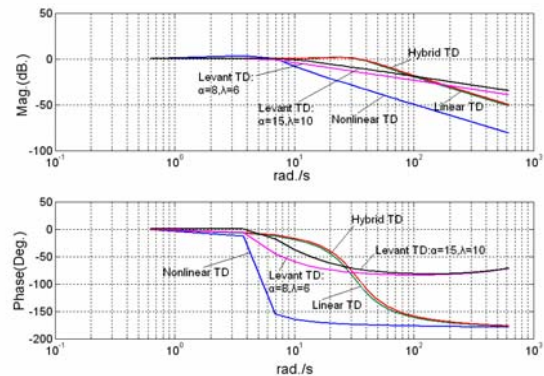

Fig. 10  Frequency characteristics

In figure 10, we can see that the hybrid differentiator has excellent tracking effect and strong ability of restraining disturbances.

Figure 11 shows the frequency characteristics of the hybrid differentiator for different $1/\varepsilon = R$ with $a_0 = 0.1, a_1 = 0.015$, $b_0 = 0.3, b_1 = 0.015$, and $\alpha_1 = \alpha_2 = 0.6$. We can conclude that the larger $R$ is, the higher frequency signal hybrid differentiator can track.

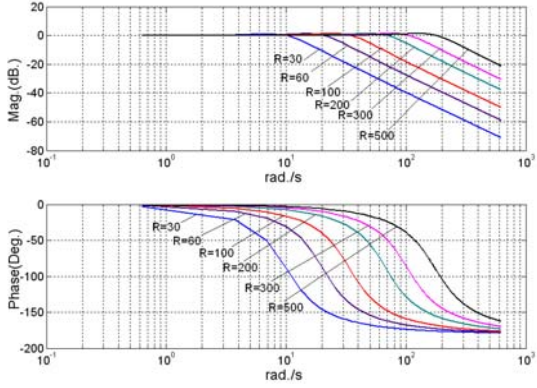

Fig. 11  Frequency characteristics of the hybrid differentiator with different $1/\varepsilon = R$

## 5. Estimation of uncertainty with noise

If the there exist noises in the uncertain system, the hybrid differentiator can estimate the uncertain. We consider the uncertain system as:

$$\dot{x} = -x + u + \delta$$
$$y = x + \xi$$

where $x \in R$ is the state, $u$ is the control input, $\delta$ is the uncertainty, and $\xi$ is the white noise. We use a hybrid differentiator to estimate $\delta$, i.e.,

$$\delta \approx \hat{\dot{x}} + x - u$$

Suppose $u = 0.1\sin t$, and $\delta = \cos t$
Hybrid differentiator:

$$\dot{x}_1 = x_2$$
$$\dot{x}_2 = -0.05 \times 45^2 (x_1 - y_{uncertain})$$
$$\quad - 0.015 \times 45^2 sig(x_1 - y_{uncertain})^{0.6}$$
$$\quad - 0.3 \times 45 x_2 - 0.015 \times 45^2 sig\left(\frac{x_2}{45}\right)^{0.6}$$
$$y = x_2$$

The simulink model is:

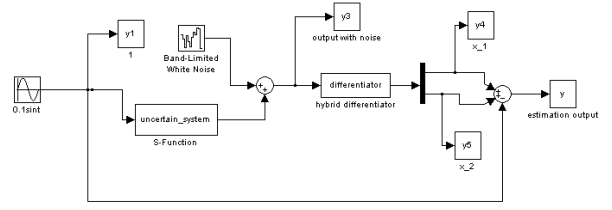

Fig. 12  Simulink model of uncertainty estimation with noise

The output $y$ of uncertain system with noises:

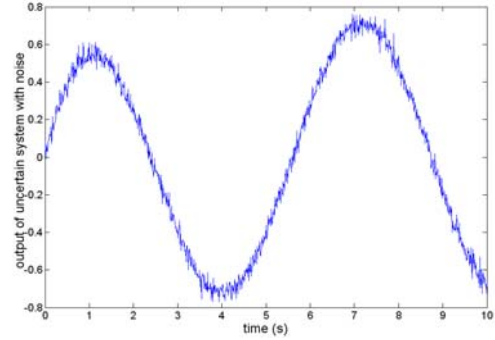

Fig. 13  Output of uncertain system with noise

Through hybrid differentiator, the estimation of uncertainty is in the following figure:

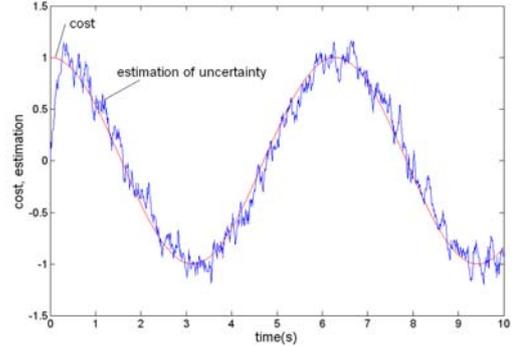

Fig. 14  Estimation of uncertainty

From Fig. 14, we can see that the estimation of uncertainty by differentiator sufficiently approaches the true one in spite of the existence of noises.

## 6. Conclusion

The describing function method is used to analyze the characteristics and parameters selection of differentiators. The parameters of hybrid differentiator can be regulated to estimate the derivative of signal and restrain noises effectively.

Acknowledgement
This paper is supported by National Nature Science Foundation of China (60774008).